\documentclass[aps,prb,twocolumn,groupedaddress,hyperref]{revtex4}
\usepackage{amsmath}
\usepackage{graphicx}
\usepackage{amsfonts}
\usepackage{amssymb}
\usepackage{float}

\providecommand{\U}[1]{\protect\rule{.1in}{.1in}}

\begin{document}

\title{Charge nonconservation of molecular devices in the presence of a
nonlocal potential}

\begin{abstract}
In the presence of a nonlocal potential in molecular device systems,
generally the charge conservation cannot be satisfied, and in literatures
the modifications of the conventional definition of current were given to
solve this problem. We demonstrate that, however, the nonconservation is not
due to the invalidation of the conventional definition of current, but
originates respectively from the improper approximations to
electron-electron interactions and the inappropriate definition of current
using pseudo wave functions in pseudopotential implementations. In this
work, we propose a nonlocal-potential formulation of the interactions to
fulfill the charge conservation and also give a discussion about the
calculation of current when the pseudopotential is involved. As an example
of application of our formulation, we further present the calculated results
of a double-barrier model.
\end{abstract}

\author{L. Q. Lai, J. Chen, Q. H. Liu, Y. B. Yu}
\email{apybyu@hnu.edu.cn}
\affiliation{School of Physics and Electronics, Hunan University, Changsha 410082, China}
\keywords{Current density; Nonlocal potential; Nonconservation; Interaction}
\pacs{}
\date{\today }
\maketitle

\section{Introduction}

Over the past few decades, investigations on transport properties of
mesoscopic systems and nanostructures have been extensively reported both on
experimental advances\cite{1,2,3,4,5,6,7,8,9,10,11,12,13} and theoretical
explorations.\cite{14,15,16,17,18,19,20,21,22,23,24,25,26,27} It is
widely acknowledged that these functional devices can be constituted by
ultrasmall conjugated molecules, single-layer or multi-layer nanotubes, bulk
organic molecules, \textit{etc.}, and plenty of interesting phenomena such
as molecular field effects,\cite{1} Coulomb blockade,\cite{2} negative
differential resistance\cite{3} and conductance switching effects\cite{4}
have been revealed, which exhibit fundamental significance and potential
microelectronic applications. In most of the works, considerable research
efforts are focused on current-voltage ($I$-$V$) characteristics as the $I$-$%
V$ profiles provide opportunities for a deeper understanding of, \textit{e.g.%
}, the basic mechanism and structure properties, as well as promising
guidance for future molecular nanoelectronics designs and manipulations.

On the theoretical side, calculations for the $I$-$V$ characteristics of
molecular device systems are mostly performed by employing the
self-consistent field (SCF) theory or nonequilibrium Green's functions
combined with density functional theory (NEGF-DFT),\cite{18,19,20} and the
widely used DFT calculations at present can be vested in the SCF method. In
comparison with conventional SCF,\cite{28,29,30} in addition to
self-consistent Hartree potential DFT introduces the exchange-correlation
potential, which is nonlocal if one wants to go beyond the local density
approximation.\cite{31} It has been shown that if the Hamiltonian $H$\
includes a general nonlocal potential $V(\mathbf{r},\mathbf{r}^{\prime })$,
an extra term naturally appears in the continuity equation and the charge
conservation will not be fulfilled.\cite{24,25,32,33} Thus the calculations
may give very incorrect results, even nonphysical results.

To resolve the problem, Li \textit{et al.}\cite{32} proposed a scheme to
modify the conventional definition of current density to include the
additional current induced by the nonlocal potential $V(\mathbf{r},\mathbf{r}%
^{\prime })$, and therefore yield the charge conservation in a
computationally efficacious way.\ However, either local or nonlocal
exchange-correlation potential stems from the approximation to
electron-electron interactions, and according to the conventional definition
of current density we will demonstrate fundamentally that with the
Hamiltonian including the exact electron-electron interactions the extra
term does not appear and the conservation can be precisely satisfied. Hence
the problem of charge nonconservation coming from the nonlocal
exchange-correlation potential should not be settled by redefining the
current density. Instead, it has to be resolved by finding a reasonable
nonlocal-potential approximation to the electron-electron interactions to
eliminate the extra term. On the other hand, norm-conserving
pseudopotentials \cite{34} are generally utilized to reduce the size of
plane-wave basis sets in first-principles calculations, which is another
origin of the nonlocal potential.\cite{35} Nevertheless, the pseudopotential
implementations give the pseudo wave functions, while in the continuity
equation $\partial _{t}\rho +\nabla \cdot \mathbf{J}=0$\ the charge density $%
\rho $ and the current density $\mathbf{J}$ should be calculated by using
the true wave functions instead of the pseudo ones. Therefore, when the
pseudopotential method is employed, the extra term appearing in the
continuity equation is due to the improper utilization of the pseudo wave
functions for calculating the current, and the problem of charge
nonconservation coming from the nonlocal pseudopotential should not be
resolved by redefining the current density either. From a fundamental point
of view, the continuity equation is a criterion regardless of any
approximations brought in as long as the particles of the system are
conserved, which can be easily proved with the original Hamiltonian. It is
the purpose of this work to investigate the above problems.

The paper is organized as follows. In Sec. II, the origins of the nonlocal
potential and the consequent issues of charge nonconservation are discussed.
We lay special emphasis on the nonlocal exchange-correlation potential from
the starting point of second quantization, and subsequently demonstrate that
in DFT calculations the nonconservation is caused by the inappropriate
definition of current using pseudo wave functions rather than the
introduction of the pseudopotentials. As an example, the currents of a
double-barrier model are numerically calculated in Sec. III to confirm our
theoretical formulation. Section IV gives the conclusions and discussions.

\section{Theoretical Formulation}

In the context of first-principles calculations, if one uses true wave
functions $\psi \left( \mathbf{r}\right) $ of the system from the very
beginning throughout the processes, the continuity equation $\partial
_{t}\rho _{c}+\nabla \cdot \mathbf{J}_{c}=0$ can be easily realized
according to the Sch\"{o}rdinger equation, where $\rho _{c}=\left\vert \psi
\left( \mathbf{r}\right) \right\vert ^{2}$ is the conventional electron
density and $\mathbf{J}_{c}$ is the conventional current density in the
absence of magnetic field with the definition as%
\begin{equation}
\mathbf{J}_{c}=\frac{-i\hbar }{2m}\left[ \psi ^{\ast }\left( \mathbf{r}%
\right) \nabla \psi \left( \mathbf{r}\right) -\psi (\mathbf{r})\nabla \psi
^{\ast }(\mathbf{r})\right] .  \label{1}
\end{equation}%
One of the origins of the charge nonconservation comes from the
approximation to electron-electron interactions, \textit{i.e.}, some
improper exchange-correlation potentials are introduced. We will show that
according to the conventional definition of current density, the
conservation is still satisfied in the presence of the interactions, but
generally can be violated by introducing the nonlocal exchange-correlation
potentials. To see this, the simplest case of Hamiltonian of a finite
many-electron system is considered, in which the electron-electron
interactions are not taken into account firstly so that the second quantized
nonrelativistic Hamiltonian (a quantity with a caret symbol denotes an
operator) is
\begin{equation}
\hat{H}_{0}=\hat{T}_{s}+\hat{U}_{ex},  \label{2}
\end{equation}%
where in terms of the field operators $\hat{\Psi}^{\dag }\left( \mathbf{r}%
,t\right) $ and $\hat{\Psi}\left( \mathbf{r},t\right) $, the single-particle
kinetic energy operator and the external potential operator can be written
respectively as\cite{36,37}%
\begin{equation}
\hat{T}_{s}=\frac{-\hbar ^{2}}{2m}\int d\mathbf{r}\hat{\Psi}^{\dag }\left(
\mathbf{r},t\right) \nabla ^{2}\hat{\Psi}\left( \mathbf{r},t\right) ,
\label{3}
\end{equation}%
and
\begin{equation}
\hat{U}_{ex}=\int d\mathbf{r}\hat{\Psi}^{\dag }\left( \mathbf{r},t\right)
v\left( \mathbf{r},t\right) \hat{\Psi}\left( \mathbf{r},t\right) .  \label{4}
\end{equation}%
The density operator can be defined as $\hat{n}(\mathbf{r},t)=$ $\hat{\Psi}%
^{\dag }\left( \mathbf{r},t\right) \hat{\Psi}\left( \mathbf{r},t\right) $
and the current density operator as the conventional form is
\begin{equation}
\hat{\mathbf{J}}_{c}(\mathbf{r},t)=\frac{-i\hbar }{2m}\left[ \hat{\Psi}%
^{\dag }\left( \mathbf{r},t\right) \nabla \hat{\Psi}\left( \mathbf{r}%
,t\right) -\left[ \nabla \hat{\Psi}^{\dag }\left( \mathbf{r},t\right) \right]
\hat{\Psi}\left( \mathbf{r},t\right) \right] ,  \label{5}
\end{equation}%
by means of the Heisenberg's equation and the anticommutation relation that
we obtain%
\begin{align}
\frac{\partial \hat{n}\left( \mathbf{r},t\right) }{\partial t}=& \frac{1}{%
i\hbar }\left[ \hat{n}\left( \mathbf{r},t\right) ,\hat{H}_{0}\right]   \notag
\\
=& \frac{i\hbar }{2m}\left[ \hat{\Psi}^{\dag }\left( \mathbf{r},t\right)
\nabla ^{2}\hat{\Psi}\left( \mathbf{r},t\right) \right.   \notag \\
& \text{ \ \ }\left. -\left[ \nabla ^{2}\hat{\Psi}^{\dag }\left( \mathbf{r}%
,t\right) \right] \hat{\Psi}\left( \mathbf{r},t\right) \right]   \notag \\
=& -\nabla \cdot \hat{\mathbf{J}}_{c}\left( \mathbf{r},t\right) ,  \label{6}
\end{align}%
where the charge conservation is accomplished, as expected. The
corresponding Hamiltonian including the interactions can be written as
\begin{equation}
\hat{H}_{w}=\hat{T}_{s}+\hat{U}_{ex}+\hat{W},  \label{7}
\end{equation}%
where
\begin{equation}
\hat{W}=\int d\mathbf{r}^{\prime \prime }d\mathbf{r}^{\prime }w\left(
\left\vert \mathbf{r}^{\prime }-\mathbf{r}^{\prime \prime }\right\vert
\right) \hat{\Psi}^{\dag }\left( \mathbf{r}^{\prime }\right) \hat{\Psi}%
^{\dag }\left( \mathbf{r}^{\prime \prime }\right) \hat{\Psi}\left( \mathbf{r}%
^{\prime \prime }\right) \hat{\Psi}\left( \mathbf{r}^{\prime }\right) .
\label{8}
\end{equation}%
Since the foregoing equations present the continuity with respect to $\hat{T}%
_{s}$ and $\hat{U}_{ex}$, only interaction operator is taken into account
hereafter, \textit{i.e.},
\begin{align}
& \left[ \hat{n}\left( \mathbf{r},t\right) ,\hat{W}\right]   \notag \\
& =\left[ \hat{\Psi}^{\dag }\left( \mathbf{r},t\right) \hat{\Psi}\left(
\mathbf{r},t\right) ,\right. \int d\mathbf{r}^{\prime \prime }d\mathbf{r}%
^{\prime }w\left( \left\vert \mathbf{r}^{\prime }-\mathbf{r}^{\prime \prime
}\right\vert \right) \hat{\Psi}^{\dag }\left( \mathbf{r}^{\prime },t\right)
\notag \\
& \text{ \ \ }\left. \times \hat{\Psi}^{\dag }\left( \mathbf{r}^{\prime
\prime },t\right) \hat{\Psi}\left( \mathbf{r}^{\prime \prime },t\right) \hat{%
\Psi}\left( \mathbf{r}^{\prime },t\right) \right]   \notag \\
& =\int d\mathbf{r}^{\prime \prime }d\mathbf{r}^{\prime }w\left( \left\vert
\mathbf{r}^{\prime }-\mathbf{r}^{\prime \prime }\right\vert \right)   \notag
\\
& \text{ \ \ }\times \left[ \hat{\Psi}^{\dag }\left( \mathbf{r},t\right)
\delta \left( \mathbf{r}-\mathbf{r}^{\prime }\right) \hat{\Psi}^{\dag
}\left( \mathbf{r}^{\prime \prime },t\right) \hat{\Psi}\left( \mathbf{r}%
^{\prime \prime },t\right) \hat{\Psi}\left( \mathbf{r}^{\prime },t\right)
\right.   \notag \\
& \text{ \ \ }-\left. \hat{\Psi}^{\dag }\left( \mathbf{r},t\right) \hat{\Psi}%
^{\dag }\left( \mathbf{r}^{\prime },t\right) \delta \left( \mathbf{r}-%
\mathbf{r}^{\prime \prime }\right) \hat{\Psi}\left( \mathbf{r}^{\prime
\prime },t\right) \hat{\Psi}\left( \mathbf{r}^{\prime },t\right) \right.
\notag \\
& \text{ \ \ }+\left. \hat{\Psi}^{\dag }\left( \mathbf{r}^{\prime },t\right)
\hat{\Psi}^{\dag }\left( \mathbf{r}^{\prime \prime },t\right) \delta \left(
\mathbf{r}-\mathbf{r}^{\prime \prime }\right) \hat{\Psi}\left( \mathbf{r}%
^{\prime },t\right) \hat{\Psi}\left( \mathbf{r},t\right) \right.   \notag \\
& \text{ \ \ }-\left. \hat{\Psi}^{\dag }\left( \mathbf{r}^{\prime },t\right)
\hat{\Psi}^{\dag }\left( \mathbf{r}^{\prime \prime },t\right) \hat{\Psi}%
\left( \mathbf{r}^{\prime \prime },t\right) \delta \left( \mathbf{r}-\mathbf{%
r}^{\prime }\right) \hat{\Psi}\left( \mathbf{r},t\right) \right]   \notag \\
& =0,  \label{9}
\end{align}%
indicating that the density operator $\hat{n}$ is commutable with the
interaction Hamiltonian $\hat{H}_{w}$, and hence the continuity equation (\ref{6}%
) is still satisfied when the interaction is involved.
In fact, the conventional definition of current and particle densities
are widely used in time-dependent current-density functional theory
(TDCDFT), and the continuity equation is frequently utilized as a constraint
condition between the two densities. \cite{45,45-1,45-2,45-3}

Next, we replace the electron-electron interaction $\hat{W}$\ by a local
potential and a nonlocal exchange-correlation potential. Since the local
potential can be absorbed by the external-field potential $\hat{U}_{ex}$,
and the interaction $\hat{W}$ can be replaced only by nonlocal
exchange-correlation potential $\hat{V}_{xc}$, thus the Hamiltonian $\hat{H}%
_{w}$ is approximated (see Appendix A) as
\begin{equation}
\hat{H}_{w}=\hat{H}_{0}+\hat{V}_{xc},  \label{10}
\end{equation}%
where
\begin{equation}
\hat{V}_{xc}=\int d\mathbf{r}d\mathbf{r}^{\prime }V_{xc}(\mathbf{r}^{\prime
},\mathbf{r})\hat{\Psi}^{\dag }(\mathbf{r}^{\prime },t)\hat{\Psi}(\mathbf{r}%
,t).  \label{11}
\end{equation}%
After some calculations we reach
\begin{align}
& \left[ \hat{n}(\mathbf{r},t),\hat{V}_{xc}\right]  \notag \\
& =\left[ \hat{\Psi}^{\dag }\left( \mathbf{r}\right) \hat{\Psi}\left(
\mathbf{r}\right) ,\int d\mathbf{r}^{\prime \prime }d\mathbf{r}^{\prime
}V_{xc}(\mathbf{r}^{\prime \prime },\mathbf{r}^{\prime })\hat{\Psi}^{\dag }(%
\mathbf{r}^{\prime \prime },t)\hat{\Psi}(\mathbf{r}^{\prime },t)\right]
\notag \\
& =\int d\mathbf{r}^{\prime \prime }d\mathbf{r}^{\prime }\hat{\Psi}^{\dag
}\left( \mathbf{r},t\right) V_{xc}(\mathbf{r}^{\prime \prime },\mathbf{r}%
^{\prime })\delta \left( \mathbf{r}-\mathbf{r}^{\prime \prime }\right) \hat{%
\Psi}(\mathbf{r}^{\prime },t)  \notag \\
& \text{ \ \ }-\int d\mathbf{r}^{\prime \prime }d\mathbf{r}^{\prime }V_{xc}(%
\mathbf{r}^{\prime \prime },\mathbf{r})\hat{\Psi}^{\dag }(\mathbf{r}^{\prime
\prime },t)\delta \left( \mathbf{r}-\mathbf{r}^{\prime }\right) \hat{\Psi}%
\left( \mathbf{r},t\right)  \notag \\
& =\int d\mathbf{r}^{\prime }\hat{\Psi}^{\dag }\left( \mathbf{r},t\right)
V_{xc}(\mathbf{r},\mathbf{r}^{\prime })\hat{\Psi}(\mathbf{r}^{\prime },t)
\notag \\
& \text{ \ \ }-\int d\mathbf{r}^{\prime \prime }\hat{\Psi}^{\dag }(\mathbf{r}%
^{\prime \prime },t)V_{xc}(\mathbf{r}^{\prime \prime },\mathbf{r})\hat{\Psi}%
\left( \mathbf{r},t\right)  \notag \\
& \neq 0,  \label{12}
\end{align}%
where the continuity condition in Eq. (\ref{6}) is no longer satisfied, and
it contains an extra term%
\begin{align}
\hat{n}_{nl}(\mathbf{r},t)=\frac{1}{i\hbar }& \int d\mathbf{r}^{\prime }%
\left[ \hat{\Psi}^{\dag }\left( \mathbf{r},t\right) V_{xc}(\mathbf{r},%
\mathbf{r}^{\prime })\hat{\Psi}(\mathbf{r}^{\prime },t)\right.  \notag \\
& \text{ \ \ }\left. -\hat{\Psi}^{\dag }(\mathbf{r}^{\prime },t)V_{xc}(%
\mathbf{r}^{\prime },\mathbf{r})\hat{\Psi}\left( \mathbf{r},t\right) \right]
,  \label{13}
\end{align}%
\textit{i.e.},
\begin{equation}
\frac{\partial \hat{n}\left( \mathbf{r},t\right) }{\partial t}+\nabla \cdot
\hat{\mathbf{J}}_{c}\left( \mathbf{r},t\right) =\hat{n}_{nl}(\mathbf{r},t).
\label{14}
\end{equation}

Thus far an inference can be drawn that the charge nonconservation is not
due to the invalidation of the conventional definition of current density,
but originates from the improper approximations to electron-electron
interactions, and a reasonable nonlocal potential arising from the
interactions should make the extra term be zero. This is the central
conclusion of this work, and the task of resolving the problem is to find
such a \textquotedblleft no-current\textquotedblright nonlocal potential.
Now, we take a mean-field approximation to the electron-electron
interactions and rewrite the potential energy as
\begin{align}
\hat{W}& =\int d\mathbf{r}d\mathbf{r}^{\prime }\frac{\Psi ^{\dagger }(%
\mathbf{r})\Psi ^{\dagger }(\mathbf{r}^{\prime })\Psi (\mathbf{r}^{\prime
})\Psi (\mathbf{r})}{|\mathbf{r}-\mathbf{r}^{\prime }|}  \notag \\
& \thickapprox \int d\mathbf{r}\Psi ^{\dagger }(\mathbf{r})\left[ \int d%
\mathbf{r}^{\prime }\frac{\langle \Psi ^{\dagger }(\mathbf{r}^{\prime })\Psi
(\mathbf{r}^{\prime })\rangle }{|\mathbf{r}-\mathbf{r}^{\prime }|}\right]
\Psi (\mathbf{r})  \notag \\
& \text{ \ \ }-\int d\mathbf{r}d\mathbf{r}^{\prime }\Psi ^{\dagger }(\mathbf{%
r})\frac{\langle \Psi ^{\dagger }(\mathbf{r}^{\prime })\Psi (\mathbf{r}%
)\rangle }{|\mathbf{r}-\mathbf{r}^{\prime }|}\Psi (\mathbf{r}^{\prime })
\notag \\
& =\int d\mathbf{r}\Psi ^{\dagger }(\mathbf{r})U_{H}(\mathbf{r})\Psi (%
\mathbf{r})  \notag \\
& \text{ \ \ }+\int d\mathbf{r}d\mathbf{r}^{\prime }\Psi ^{\dagger }(\mathbf{%
r})U_{xc}(\mathbf{r}^{\prime },\mathbf{r})\Psi (\mathbf{r}^{\prime }),
\label{15}
\end{align}%
where $\langle ...\rangle $\ represents the ensemble statistical average,
\begin{equation}
U_{H}(\mathbf{r})=\int d\mathbf{r}^{\prime }\frac{\langle \Psi ^{\dagger }(%
\mathbf{r}^{\prime })\Psi (\mathbf{r}^{\prime })\rangle }{|\mathbf{r-r}%
^{\prime }|},  \label{16}
\end{equation}%
and%
\begin{equation}
U_{xc}(\mathbf{r,r}^{\prime })=-\frac{\langle \Psi ^{\dagger }(\mathbf{r}%
^{\prime })\Psi (\mathbf{r})\rangle }{|\mathbf{r-r}^{\prime }|}.  \label{17}
\end{equation}%
Therefore we have obtained our nonlocal exchange-correlation potential.
Introducing an auxiliary variable $\hat{F}$\ as%
\begin{align}
\hat{F}(\mathbf{r})=\frac{1}{i\hbar }& \int d\mathbf{r}^{\prime }\left[
U_{xc}(\mathbf{r,r}^{\prime })\Psi ^{^{\dagger }}(\mathbf{r})\Psi (\mathbf{r}%
^{\prime })\right.   \notag \\
& \text{ \ \ }\left. -U_{xc}(\mathbf{r}^{\prime }\mathbf{,r})\Psi ^{\dagger
}(\mathbf{r}^{\prime })\Psi (\mathbf{r})\right] ,  \label{18}
\end{align}%
and similarly we have%
\begin{equation}
\frac{\partial \hat{n}\left( \mathbf{r},t\right) }{\partial t}+\nabla \cdot
\hat{\mathbf{J}}_{c}\left( \mathbf{r},t\right) =\hat{F}(\mathbf{r}),
\label{19}
\end{equation}%
where an extra term is still contained in the meaning of operator. However,
after taking the ensemble statistical average \cite{38} of $\hat{F}(\mathbf{r%
}),$ we can obtain%
\begin{align}
\langle \hat{F}(\mathbf{r})\rangle =\frac{1}{i\hbar }& \int d\mathbf{r}%
^{\prime }\left[ U_{xc}(\mathbf{r,r}^{\prime })\langle \Psi ^{\dagger }(%
\mathbf{r})\Psi (\mathbf{r}^{\prime })\rangle \right.   \notag \\
& \text{ \ \ }\left. -U_{xc}(\mathbf{r}^{\prime }\mathbf{,r})\langle \Psi
^{\dagger }(\mathbf{r}^{\prime })\Psi (\mathbf{r})\rangle \right]   \notag \\
=\frac{1}{i\hbar }& \int d\mathbf{r}^{\prime }\left[ \frac{\langle \Psi
^{\dagger }(\mathbf{r}^{\prime })\Psi (\mathbf{r})\rangle }{|\mathbf{r-r}%
^{\prime }|}\langle \Psi ^{\dagger }(\mathbf{r})\Psi (\mathbf{r}^{\prime
})\rangle \right.   \notag \\
& \text{ \ \ }\left. -\frac{\langle \Psi ^{\dagger }(\mathbf{r})\Psi (%
\mathbf{r}^{\prime })\rangle }{|\mathbf{r-r}^{\prime }|}\langle \Psi
^{\dagger }(\mathbf{r}^{\prime })\Psi (\mathbf{r})\rangle \right]   \notag \\
& =0,  \label{20}
\end{align}%
\textit{i.e.},
\begin{equation}
\frac{\partial \langle \hat{n}\left( \mathbf{r},t\right) \rangle }{\partial t%
}+\nabla \cdot \langle \hat{\mathbf{J}}_{c}\left( \mathbf{r},t\right)
\rangle =0,  \label{21}
\end{equation}%
which indicates that in the meaning of statistical average the conservation
can be again fulfilled. This result is quite satisfactory, because an
observable quantity is a statistical-average one. One may find that the
nonlocal exchange-correlation potential in Eq. (\ref{17}) is similar to the
Fock term of the Hartree-Fock approximation. In the case of zero
temperature, this proposed nonlocal exchange-correlation potential is just
the Fock term, and in other cases they are different. Normal Hartree-Fock
approximation is only applicable to the ground state, while Eq. (\ref{17})
can be used in nonequilibrium states.

Another origin of the nonconservation is the inappropriate definition of
current density in pseudopotential implementations. To conserve the current,
one should calculate the current density $\mathbf{J}$ by using the true wave
function $\psi \left( \mathbf{r}\right)$ rather than the generally obtained
pseudo wave function $\phi \left( \mathbf{r}\right)$ (see Appendix B). Note
that in practical calculations, the strict pseudopotential would be somewhat
difficult to be used due to the explicit energy dependence, and approximate
model pseudopotentials are often introduced instead. As shown in Eq. (\ref%
{27}), the nonlocal part of the pseudopotential comes from the core wave
functions, which is confined in a small core region surrounding the atom,
and the true and pseudo wave functions are identical outside this small
region.\cite{34,39} Thus the additional current induced by the nonlocal
pseudopotential is localized in the core of the atom, and its contribution
to the transport current can be neglected.

\begin{figure}[tbp]
\centering
\includegraphics[height=30mm, width=65mm]{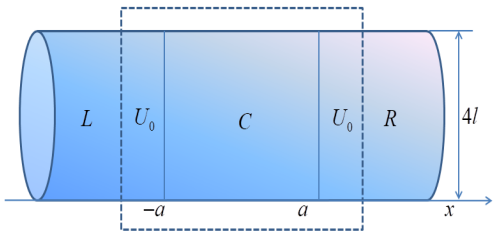}
\caption{Schematic view of the transversely confined double-barrier model.
The dashed-line box \textit{C }is a hypothetical region including the core
device and part of two ideal leads, and \textit{L/R }symbolize the rest of
the leads.}
\label{fig:1}
\end{figure}

\section{Numerical Implementation}

To illustrate the theoretical formulation proposed above, we consider a
quasi-one-dimensional double-barrier model confined transversely to simulate
the device system, as shown in Fig. 1. The nonlocal potential is placed only
in the region between $-a$ and $a$, and the barriers can be regarded as part
of two ideal leads without nonlocal potentials.

\begin{figure}[h]
\centering
\includegraphics[height=55mm, width=80mm]{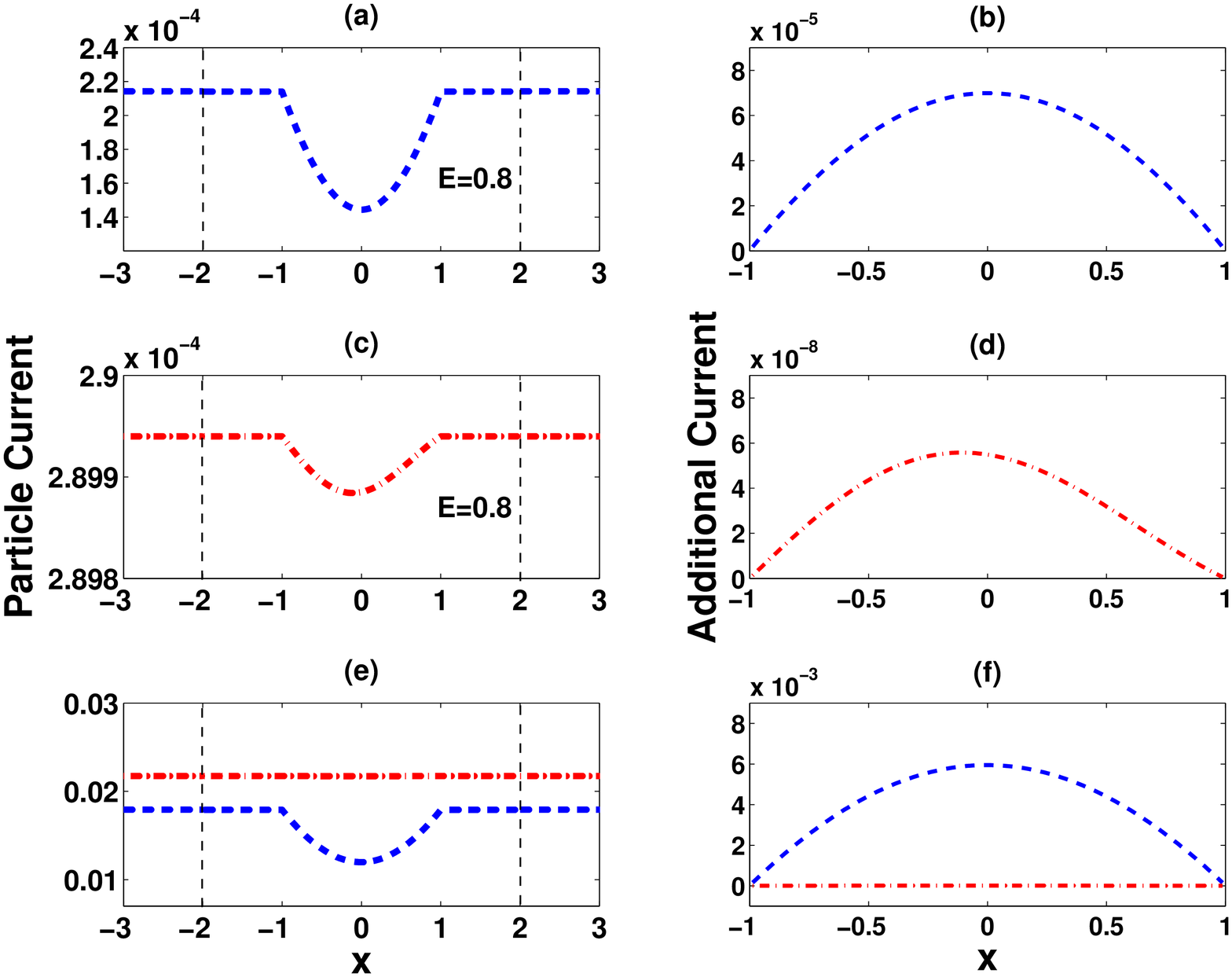}
\caption{Conventional particle current $I_{c}$ (left column) and additional
current $I_{ad}$ (right column) are along the $x$-direction. (a) and (b) are
the currents of single-particle with energy $E=0.8$ for the model nonlocal
potential $V(x,x^{\prime })$; (c) and (d) are the currents of
single-particle with energy $E=0.8$ for our proposed nonlocal potential; (e)
and (f) are the total currents of all the particles below the Fermi energy $%
E_{F}$ calculated with our proposed nonlocal potential (red dash-dot line) and
the model nonlocal potential (blue dash line). The vertical dash lines at $x=-2$
and $x=2$ mark the contact regions. Here, $\protect\lambda =0.009$, $\protect%
\eta =0.01$, and the width of the barriers is $1.0$.}
\label{fig:2}
\end{figure}

We imagine a central region \textit{C} (the dashed-line box) that encloses
both the core device and the barriers, and the total Hamiltonian for this
model can be written as%
\begin{equation}
\hat{H}_{tot}=\hat{H}_{0}+\hat{V}_{H}+\hat{V}_{xc},  \label{31}
\end{equation}%
where $\hat{H}_{0}=\frac{-\hbar ^{2}}{2m}\nabla ^{2}+U_{0}$ with $U_{0}$ the
local potential within the barriers, $\hat{V}_{H}$ is the Hartree potential
and $\hat{V}_{xc}$ is the exchange-correlation potential: $\langle \mathbf{r|%
}\hat{V}_{xc}|\psi \rangle =\int V_{xc}\left( \mathbf{r,r}^{\prime }\right)
\psi (\mathbf{r}^{\prime })d\mathbf{r}^{\prime }$. We simplify the potential
in Eq. (\ref{16}) and Eq. (\ref{17}) to the form of one dimension as%
\begin{equation}
V_{H}\left( x\right) =\int {\sum\limits_{k}}v\left( x,x^{\prime }\right)
|\psi _{k}\left( x^{\prime }\right) |^{2}dx^{\prime },  \label{32}
\end{equation}%
and
\begin{equation}
V_{xc}\left( x,x^{\prime }\right) =-{\sum\limits_{k}}v\left( x,x^{\prime
}\right) \psi _{k}^{\ast }\left( x\right) \psi _{k}\left( x^{\prime }\right)
,  \label{33}
\end{equation}%
where $\psi_{k} \left(x\right)$ is the longitudinal wave function, and
\begin{align}
v\left( x,x^{\prime }\right) & =\int \rho _{1}d\rho _{1}\int \rho _{2}d\rho
_{2}\int \frac{d\varphi }{2\pi }  \notag \\
& \text{ \ \ }\times \frac{\varphi ^{2}\left( \rho _{1}\right) \varphi
^{2}\left( \rho _{2}\right) }{\sqrt{\rho _{1}^{2}+\rho _{2}^{2}-2\rho
_{1}\rho _{2}\cos \varphi +\left( x-x^{\prime }\right) ^{2}}},  \label{34}
\end{align}%
which is averaged over the transverse wave functions.\cite{40} A single
transverse wave function $\varphi \left( \rho \right) =1/\sqrt{2\pi l^{2}}%
\exp \left( -\rho ^{2}/4l^{2}\right) $ is chosen universally, and a
transverse radius $2l$ characterizes the size of the confinement. To
proceed, we first solve the energy eigenequation of the double-barrier model
with only local potential $U_{0}$ and numerically calculate the wave
functions $\psi \left( x\right) $, thus the corresponding Hartree potential $%
V_{H}$ and exchange-correlation potential $V_{xc}$ of the central region can
be constructed from the wave functions\emph{\ }by using Eq. (\ref{32}) and
Eq. (\ref{33}). Then we return to the calculation of the wave functions $%
\psi \left( x\right)$, and self-consistent calculations are performed using
the iterative procedure in the numerical implementation. In general, once
all the wave functions with eigenenergies below the Fermi energy $E_{F}$ are
known, we can obtain the current density at all points. When the system is
under applied bias, the Fermi levels in the left and right leads are taken
respectively as $E_{F}+eV$\ and $E_{F},$\ where $E_{F}$\ is the Fermi level
in the equilibrium state and $V$\ is the bias, \textit{i.e.}, the sums in
Eq. (\ref{32}) and Eq. (\ref{33}) are up to $E_{F}+eV$\ and $E_{F}$
respectively for left-incident and right-incident modes of the functions.
Energy unit $\hbar ^{2}/2ma^{2}$ is used throughout the calculations as a
reduced coefficient, where $a$ is the width of the barrier, $m$ is the mass
of the electron. The Fermi energy here is set as $E_{F}=1.0$, and the
magnitude of the barriers is fixed to be $U_{0}=6.0$. In addition, all the
calculations are preformed in low-temperature limit, \textit{i.e.}, the
temperature $T=0$.

\begin{figure}[tbp]
\centering
\includegraphics[height=55mm, width=80mm]{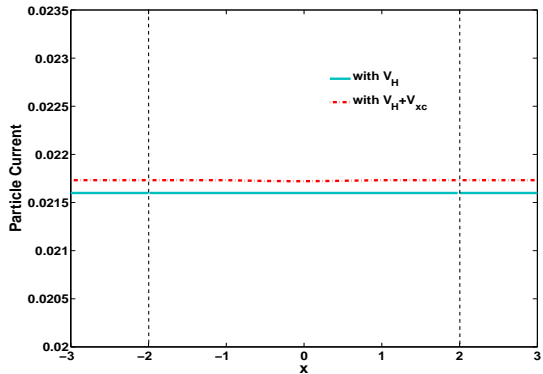}
\caption{Comparison of the particle currents for each region. The case that
involves the proposed nonlocal exchange-correlation potential $V_{xc}$
(dash-dot line) is somewhat a correction to the one that only considers the
Hartree potential $V_{H}$ (solid line).}
\label{fig:3}
\end{figure}

We first study the influences of a model nonlocal potential and our proposed
potential in Eq. (\ref{17}) on the particle currents, respectively. The
nonlocal potential is nonzero when $x$ is from $-1$ to $1$, and the barriers
are located in $\left[ -2,-1\right] $ and $\left[ 1,2\right] $. The model
nonlocal potential here is chosen as $V\left( x,x^{\prime}\right)
=\lambda\exp\left[ \eta\left( x-x^{\prime}\right) ^{2}\right] $, where $%
\lambda$ and $\eta$ are two independent coefficients. In Fig. 2, we present
the calculated currents of left-incoming electrons with a fixed energy $E=0.8
$ and below the Fermi energy $E_{F}$, respectively. Here, both the
conventional particle current and additional current are given, and the
additional currents coming from nonlocal potential $V(x,x^{\prime})$ are
calculated by
\begin{align}
I_{ad}(x) =\frac{-1}{i\hbar }\int_{-1}^{x}dx^{\prime
\prime}\int_{-1}^{1}dx^{\prime }&\sum\limits_{k}\left[\psi _{k}^{\ast
}\left( x^{\prime \prime }\right) V(x^{\prime \prime },x^{\prime
})\psi_{k}\left( x^{\prime}\right) \right.  \notag \\
& \text{ \ \ } \left. -\psi _{k}^{\ast }\left( x^{\prime }\right)
V(x^{\prime },x^{\prime\prime })\psi _{k}(x^{\prime \prime })\right].  \notag
\end{align}
From Fig. 2(a) to Fig. 2(d), one can see that for single-particle with the
fixed energy the left and right lead currents hold to be constant, while in
the central region the currents turn out to be varying with $x$, and the
additional currents from the nonlocal potential are nonzero for both the
model potential and our proposed potential. It means that in the case of 
single-particle, the lead currents and the central current calculated with the
proposed nonlocal potential behave similarly to that obtained by using the
model nonlocal potential, which also violates the conservation. However,
Figures 2(e) and 2(f) show that when currents of all incoming electrons
below the Fermi energy $E_{F}$ are considered, in the case of the model
nonlocal potential the conventional current still violates the continuity
condition due to the nonzero additional current, while the conventional
currents calculated with the proposed nonlocal potential are seen to be
conservative in the whole simulation region and the corresponding additional
current vanishes. Moreover, particle currents in all the three regions serve
as a minor correction to the ones with Hartree potential only (see Fig. 3).

\begin{figure}[tbp]
\centering
\includegraphics[height=55mm, width=80mm]{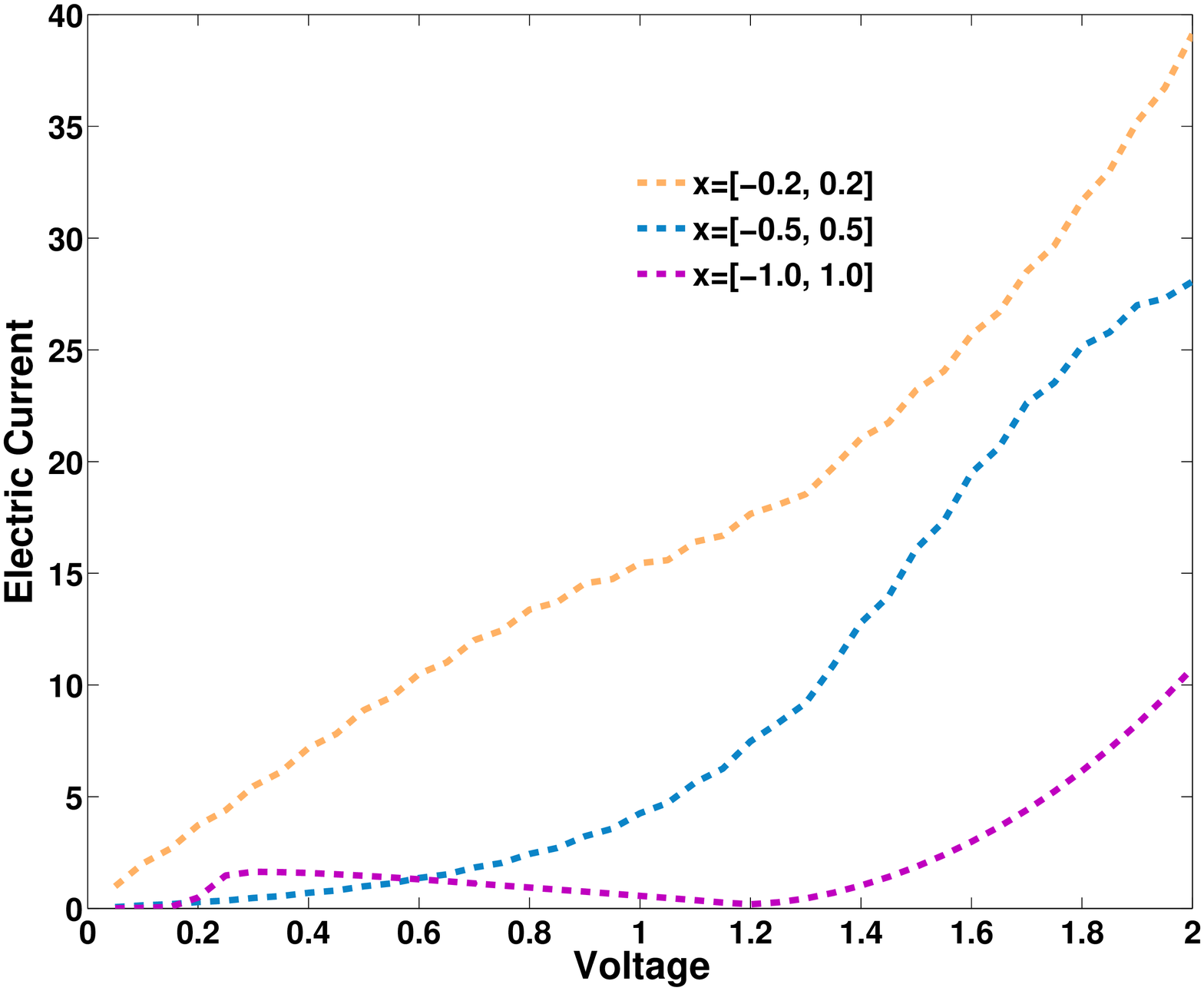}
\caption{Dependence of the electric currents on the selection of the central
region. Nonlocal potential is not included, and the width and magnitude of
the barriers are both fixed.}
\label{fig:4}
\end{figure}

\begin{figure}[tbp]
\centering
\includegraphics[height=55mm, width=80mm]{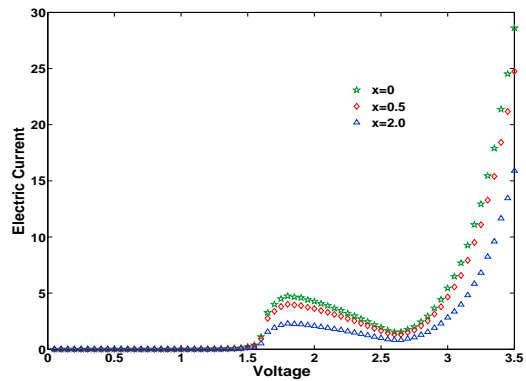}
\caption{Current properties of different sites of the system with the model
nonlocal potential $V_{nl}$. The central region is fixed to be $x=\left[ -1,1%
\right] $, and the barriers are located in $x=\left[ -2,-1\right] $ and $x=%
\left[ 1,2\right] $. Discrepancies of the currents for all three positions
enlarge with the increasing voltage.}
\label{fig:5}
\end{figure}

Figure 4 plots the dependence of the electric current on the selection of
each region. The width of the barriers are fixed as the above while the
whole central region is changeable, and the nonlocal potential is absent
therein. As we shall see, when the core region is chosen as small as $%
x=[-0.2,0.2]$, it exhibits a sublinear relation between the current $I$ and
the voltage $V$, which is similar to that of a general single-barrier model,
and the scattering effect in the central box vanishes. With the increases of
the width as $x=\left[ -0.5,0.5\right] $ and $x=\left[ -1.0,1.0\right] $,
the curves shift downwards gradually and finally become nonlinear ones,
indicating that careful considerations on relevant area selection must be
taken into actual systems to ensure the computational accuracy.

In Fig. 5, we present the $I$-$V$ characteristics of different sites along
the $x$ axis when the above model nonlocal potential $V\left(
x,x^{\prime}\right) $ is included. Currents of $x=0$, $x=0.5$ and $x=2.0$
differ mutually, as expected, which implies that currents in the central
region and the leads calculated with the model nonlocal potential cannot
meet the charge conservation. We further explore the contrastive
characteristics with our approximation. From Fig. 6, one can see that when
only Hartree potential $V_{H}$ is taken into account, currents of the three
sites are equal to each other with the increase of the voltage, while the
whole of the curve bears an upward shift when the proposed
exchange-correlation potential $V_{xc}$ is also considered, and the currents
obtained can be still conservative.

\begin{figure}[tbp]
\centering
\includegraphics[height=55mm, width=80mm]{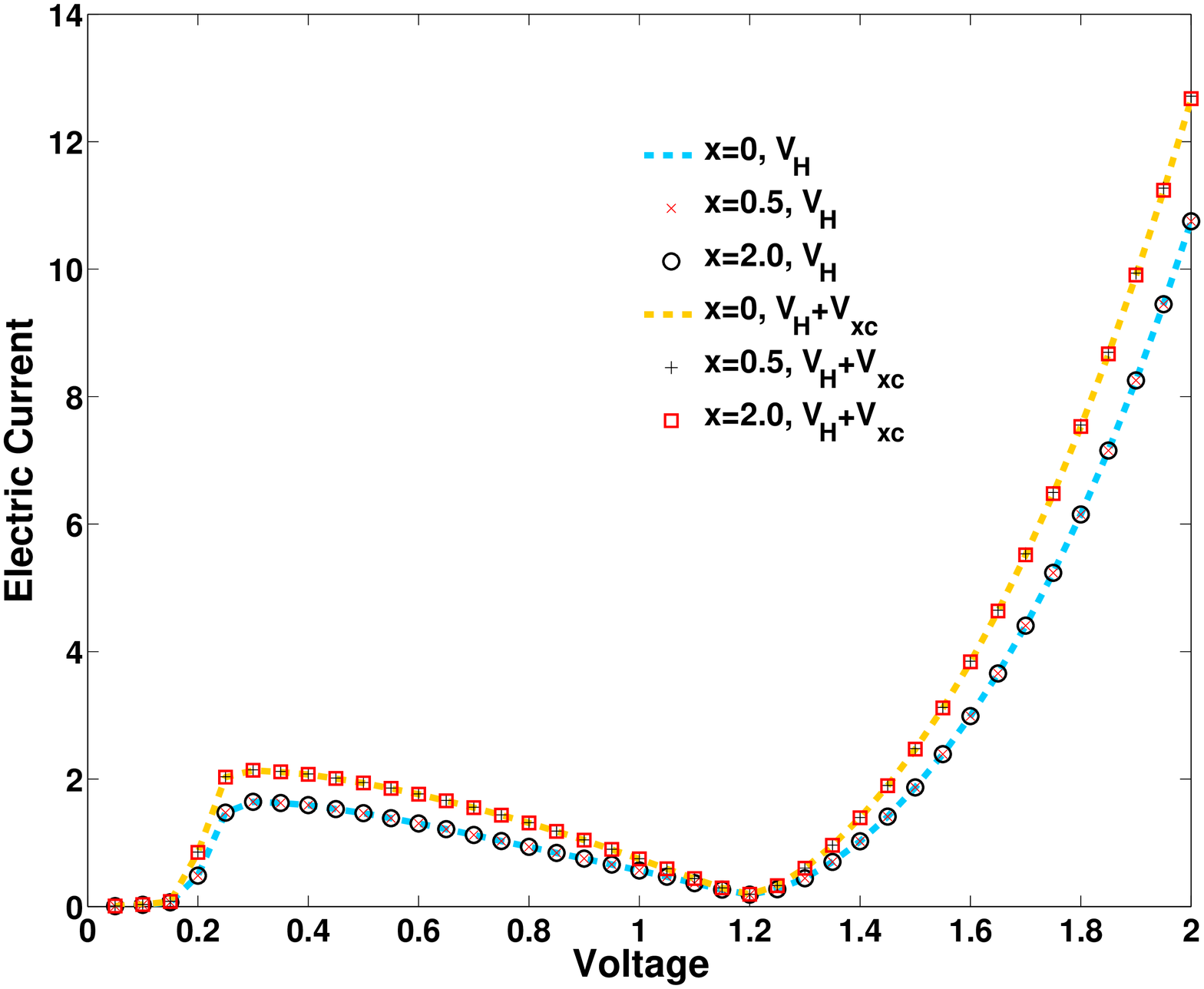}
\caption{$I$-$V$ characteristics of different sites for: (1) with only
Hartree potential $V_{H}$, and (2) with both the Hartree potential $V_{H}$
and the nonlocal exchange-correlation potential $V_{xc}$. The currents
obtained from the proposed formalism satisfy the charge conservation.}
\label{fig:6}
\end{figure}

\section{Conclusions and Discussions}

In summary, it is of great importance to give correct dynamic charge and
potential distributions of the transport systems, which is the key point for
the valid currents and prospective applications of molecular devices. Once
we are able to calculate the current strictly according to the original
Hamiltonian including electron-electron interactions, the problems of charge
nonconservation induced by nonlocal potential would not exist, and the
results are undoubtedly reasonable. However, it is impossible to obtain
rigorous solutions under such intricate interactions. We demonstrate these
issues and attest that the nonconservation stems respectively from the
improper approximations to electron-electron interactions and the
inappropriate definition of current using pseudo wave functions in
pseudopotential implementations, and propose a nonlocal-potential
formulation of the interactions to fulfill the conservation in the meaning
of statistical average, as well as give a verification about the calculation
of current in the presence of pseudopotentials. With this method, we have
also studied a double-barrier model to simulate the molecular device system,
which further confirms our formulation.

\section*{ACKNOWLEDGEMENTS}
This work is supported by National Natural Science Foundation of China under Grant No. 11675051.

\begin{appendix}
\section{}

Here, we will attempt to give the second quantized form of the nonlocal
potential. According to common methods, the wave function $\psi $ can be
expanded in terms of a complete basis set $\varphi _{m}$ as%
\begin{equation}
\psi \left( \mathbf{r},t\right) =\sum_{m}c_{m}\left( t\right) \varphi
_{m}\left( \mathbf{r}\right) ,  \label{A1}
\end{equation}%
where $c_{m}$ is a coefficient independent of $\mathbf{r}$. Substituting the
above wave function into the following Schr\"{o}dinger equation, which
includes the nonlocal potential $V_{xc}(\mathbf{r},\mathbf{r}^{\prime })$,

\begin{equation}
i\hbar \frac{\partial }{\partial t}\psi (\mathbf{r},t)=H_{0}\psi (\mathbf{r}%
,t)+\int d\mathbf{r}^{\prime }V_{xc}(\mathbf{r},\mathbf{r}^{\prime })\psi (%
\mathbf{r}^{\prime },t),  \label{A2}
\end{equation}%
we have
\begin{equation}
i\hbar \dot{c}_{m}(t)=\sum_{n}H_{0mn}c_{n}(t)+\sum_{n}V_{mn}c_{n}(t),
\label{A3}
\end{equation}%
where%
\begin{equation}
H_{0}=\frac{-\hbar ^{2}}{2m}\nabla ^{2}+v\left( \mathbf{r},t\right) ,
\label{4}
\end{equation}%
\begin{equation}
H_{0mn}=\int d\mathbf{r}d\mathbf{r}^{\prime }\varphi _{m}^{\ast }(\mathbf{r}%
)H_{0}\varphi _{n}(\mathbf{r}^{\prime }),  \label{A5}
\end{equation}%
and%
\begin{equation}
V_{mn}=\int d\mathbf{r}d\mathbf{r}^{\prime }\varphi _{m}^{\ast }(\mathbf{r}%
)V_{xc}(\mathbf{r},\mathbf{r}^{\prime })\varphi _{n}(\mathbf{r}^{\prime })
\label{A6}
\end{equation}%
is the matrix elements of $V_{xc}(\mathbf{r},\mathbf{r}^{\prime })$.
Similarly, the field operator is%
\begin{equation}
\hat{\Psi}(\mathbf{r},t)=\sum_{n}\varphi _{n}(\mathbf{r})\hat{a}_{n}(t),
\label{A7}
\end{equation}%
with
\begin{equation}
\hat{a}_{n}(t)=\int d\mathbf{r}\varphi _{n}^{\ast }(\mathbf{r},t)\hat{\Psi}(%
\mathbf{r},t),  \label{A8}
\end{equation}%
where $\hat{a}_{n}^{\dag }$ $\left( \hat{a}_{n}\right) $\ is the creation
(annihilation) operator. Therefore, we obtain the second quantized form of the
nonlocal potential as%
\begin{align}
\hat{V}_{xc}& =\sum_{mn}V_{mn}(t)\hat{a}_{m}^{\dag }(t)\hat{a}_{n}(t)
\notag \\
& =\int d\mathbf{r}d\mathbf{r}^{\prime }\sum_{mn}V_{mn}(t)\varphi _{n}^{\ast
}(\mathbf{r},t)\varphi _{m}(\mathbf{r}^{\prime },t)\hat{\Psi}^{\dag }(%
\mathbf{r}^{\prime },t)\hat{\Psi}(\mathbf{r},t)  \notag \\
& =\int d\mathbf{r}d\mathbf{r}^{\prime }V_{xc}(\mathbf{r}^{\prime },\mathbf{r%
})\hat{\Psi}^{\dag }(\mathbf{r}^{\prime },t)\hat{\Psi}(\mathbf{r},t).
\label{A9}
\end{align}

\section{}

We first discuss that in introducing the nonlocal pseudopotential, the current
calculated with pseudo wave functions does not satisfy the conservation.
Pseudopotentials were originally introduced to simplify electronic structure
calculations by adding some core functions to the true wave function $\psi
\left( \mathbf{r}\right) $ to obtain a smooth pseudo wave function\cite%
{41,42,43}%
\begin{equation}
\psi \left( \mathbf{r}\right) =\phi \left( \mathbf{r}\right)
-\sum\limits_{n}\langle \psi _{n}|\phi \rangle \psi _{n}\left( \mathbf{r}%
\right) ,  \label{22}
\end{equation}%
where $\phi \left( \mathbf{r}\right) $ is the pseudo wave function and $\psi
_{n}\left( \mathbf{r}\right) $ is the core function, and the general form of
the pseudopotential (only its nonlocal part) is%
\begin{equation}
V_{nl}=\sum\limits_{n}(E-E_{n})|\psi _{n}\rangle \langle \psi _{n}|,
\label{23}
\end{equation}%
from which we obtain%
\begin{equation}
\sum\limits_{n}|\psi _{n}\rangle \langle \psi _{n}|=\left( E-H_{0}\right)
^{-1}V_{nl}=GV_{nl},  \label{24}
\end{equation}%
where $G=\left( E-H_{0}\right) ^{-1}$ is the Green's function, and $H_{0}$
is the original Hamiltonian. The true wave function can then be denoted by%
\begin{align}
\psi \left( \mathbf{r}\right) & =\langle \mathbf{r}|\psi \rangle  \notag \\
& =\langle \mathbf{r}|\phi \rangle -\langle \mathbf{r}|GV_{nl}|\phi \rangle
\notag \\
& =\phi \left( \mathbf{r}\right) -\int d\mathbf{r}^{\prime \prime }\int d%
\mathbf{r}^{\prime }\langle \mathbf{r}|G|\mathbf{r}^{\prime }\rangle \langle
\mathbf{r}^{\prime }|V_{nl}|\mathbf{r}^{\prime \prime }\rangle \langle
\mathbf{r}^{\prime \prime }|\phi \rangle  \notag \\
& =\phi \left( \mathbf{r}\right) -\int d\mathbf{r}^{\prime }d\mathbf{r}%
^{\prime \prime }G(\mathbf{r,r}^{\prime })V_{nl}(\mathbf{r}^{\prime },%
\mathbf{r}^{\prime \prime })\phi \left( \mathbf{r}^{\prime \prime }\right) .
\label{25}
\end{align}%
Therefore, the corresponding Schr\"{o}dinger equation for the pseudo wave
function is%
\begin{equation}
H_{0}\phi \left( \mathbf{r}\right) +\int d\mathbf{r}^{\prime }V_{nl}(\mathbf{%
r,r}^{\prime })\phi \left( \mathbf{r}^{\prime }\right) =E\phi \left( \mathbf{%
r}\right) ,  \label{26}
\end{equation}%
where%
\begin{align}
V_{nl}(\mathbf{r,r}^{\prime })& =\sum\limits_{n}(E-E_{n})\psi _{n}(\mathbf{r}%
)\psi _{n}^{\ast }(\mathbf{r}^{\prime })  \notag \\
& =(E-H_{0})\sum\limits_{n}\psi _{n}(\mathbf{r})\psi _{n}^{\ast }(\mathbf{r}%
^{\prime }).  \label{27}
\end{align}

It is easy to verify that the conventionally defined current density $%
\mathbf{J}=\frac{-i\hbar }{2m}[\phi ^{\ast }\left( \mathbf{r}\right) \nabla
\phi \left( \mathbf{r}\right) -\phi \left( \mathbf{r}\right) \nabla \phi
^{\ast }\left( \mathbf{r}\right) ]$ along with the obtained pseudo electron
density $\rho _{ps}=\left\vert \phi \left( \mathbf{r}\right) \right\vert
^{2} $ do not meet the charge conservation. Nevertheless, this
nonconservation is not caused by introducing the pseudopotential, but is
that the above definition of current density cannot be used with the pseudo
wave function $\phi \left( \mathbf{r}\right) $. The correct current density
must be defined by using the true wave function $\psi \left( \mathbf{r}%
\right) $. According to the definition $\mathbf{J}_{r}=$ $\frac{-i\hbar }{2m}%
\left[ \psi ^{\ast }\left( \mathbf{r}\right) \nabla \psi \left( \mathbf{r}%
\right) -\psi (\mathbf{r})\nabla \psi ^{\ast }(\mathbf{r})\right] $, we get
\begin{equation}
\mathbf{J}_{r}=\mathbf{J}_{ps}+\mathbf{J}_{nl},  \label{28}
\end{equation}%
where%
\begin{equation}
\mathbf{J}_{ps}=\frac{-i\hbar }{2m}\left[ \phi ^{\ast }\left( \mathbf{r}%
\right) \nabla \phi \left( \mathbf{r}\right) -\phi \left( \mathbf{r}\right)
\nabla \phi ^{\ast }\left( \mathbf{r}\right) \right] ,  \label{29}
\end{equation}%
and%
\begin{align}
\mathbf{J}_{nl}=\frac{i\hbar }{2m}& \left[ \phi ^{\ast }\left( \mathbf{r}%
\right) \nabla \Pi (\mathbf{r)+}\Pi (\mathbf{r)}\nabla \phi ^{\ast }\left(
\mathbf{r}\right) \right.  \notag \\
& \text{ \ \ }\left. -\Pi ^{\ast }\left( \mathbf{r}\right) \nabla \Pi (%
\mathbf{r)-}c.c.\right] .  \label{30}
\end{align}%
Here, $\Pi (\mathbf{r)=}\int d\mathbf{r}^{\prime }d\mathbf{r}^{\prime \prime
}G(\mathbf{r,r}^{\prime })V_{nl}(\mathbf{r}^{\prime },\mathbf{r}^{\prime
\prime })\phi \left( \mathbf{r}^{\prime \prime }\right) $. On the basis of
Eq. (\ref{26}), in steady states we can easily give $\nabla \cdot \mathbf{J}%
_{ps}=-\nabla \cdot \mathbf{J}_{nl}$, which verifies the conservation with
the definition of current density in Eq. (\ref{28}).

\end{appendix}


\bigskip

\begin{thebibliography}{99}
\bibitem{1} J. Paloheimo, P. Kuivalainen, H. Stubb, E. Vuorimaa, and P.
Yli-Lahti, Appl. Phys. Lett. \textbf{56}, 1157(1990).

\bibitem{2} L. P. Kouwenhoven, A. T. Johnson, N. C. van der Vaart, C. J. P.
M. Harmans, and C. T. Foxon, Phys. Rev. Lett. \textbf{67}, 1626 (1991).

\bibitem{3} J. Chen, M. A. Reed, A. M. Rawlett, and J. M. Tour, Science
\textbf{286}, 1550 (1999).

\bibitem{4} C. P. Collier, E. W. Wong, M. Belohradsk\'{y}, F. M. Raymo, J.
F. Stoddart, P. J. Kuekes, R. S. Williams, and J. R. Heath, Science \textbf{%
285}, 391 (1999).

\bibitem{5} C. Joachim, J. K. Gimzewski, and A. Aviram, Nature \textbf{408},
541 (2000).

\bibitem{6} J. Park, A. N. Pasupathy, J. I. Goldsmith, C. Chang, Y. Yaish,
J. R. Petta, M. Rinkoski, J. P. Sethna, H. D. Abru\~{n}a, P. L. McEuen, and
D. C. Ralph, Nature \textbf{417}, 722 (2002).

\bibitem{7} N. J. Tao, Nat. Nanotechnol. \textbf{1}, 173 (2006).

\bibitem{8} S. M. Lindsay and M. A. Ratner, Adv. Mater. \textbf{19}, 23
(2007).

\bibitem{9} Y. M. Lin, K. A. Jenkins, A. Valdes-Garcia, J. P. Small, D. B.
Farmer, and P. Avouris, Nano Lett. \textbf{9}, 422 (2009).

\bibitem{10} N. Rauhut, M. Engel, M. Steiner, R. Krupke, P. Avouris, and A.
Hartschuh, ACS Nano \textbf{6}, 6416 (2012).

\bibitem{11} B. Xu and Y. Dubi, J. Phys.: Condens. Matter \textbf{27},
263202 (2015).

\bibitem{12} K. Ono, G. Giavaras, T. Tanamoto, T. Ohguro, X. Hu, and F.
Nori, Phys. Rev. Lett. \textbf{119}, 156802 (2017).

\bibitem{13} Y. Isshiki, S. Fujii, T. Nishino, and M. Kiguchi, J. Am. Chem.
Soc. \textbf{140}, 3760 (2018).

\bibitem{14} R. Landauer, IBM J. Res. Dev. \textbf{1}, 233 (1957).

\bibitem{15} M. B\"{u}ttiker, Y. Imry, R. Landauer, and S. Pinhas, Phys.
Rev. B \textbf{31}, 6207 (1985).

\bibitem{16} A. Pr\^{e}tre, H. Thomas, and M. B\"{u}ttiker, Phys. Rev. B
\textbf{54}, 8130 (1996).

\bibitem{17} A. P. Jauho, N. S. Wingreen, and Y. Meir, Phys. Rev. B \textbf{%
50}, 5528 (1994).

\bibitem{18} B. Wang, J. Wang, and H. Guo, Phys. Rev. Lett. \textbf{82}, 398
(1999).

\bibitem{19} J. Taylor, H. Guo, and J. Wang, Phys. Rev. B \textbf{63},
245407 (2001).

\bibitem{20} M. Brandbyge, J. L. Mozos, P. Ordej\'{o}n, J. Taylor, and K.
Stokbro, Phys. Rev. B \textbf{65}, 165401 (2002).

\bibitem{21} S. Kurth, G. Stefanucci, C. -O. Almbladh, A. Rubio, and E. K.
U. Gross, Phys. Rev. B \textbf{72}, 035308 (2005).

\bibitem{22} C. Y. Yam, X. Zheng, G. H. Chen, Y. Wang, T. Frauenheim, and T.
A. Niehaus, Phys. Rev. B \textbf{83}, 245448 (2011).

\bibitem{23} K. Varga, Phys. Rev. B \textbf{83}, 195130 (2011).

\bibitem{24} L. Zhang, B. Wang, and J. Wang, Phys. Rev. B \textbf{84},
115412 (2011).

\bibitem{25} K. T. Cheung, B. Fu, Z. Z. Yu, and J. Wang, Phys. Rev. B
\textbf{95}, 125422 (2017).

\bibitem{26} H. Wang and M. Thoss, J. Chem. Phys. \textbf{138}, 134704
(2013).

\bibitem{27} G. Cabra, A. Jensen, and M. Galperin, J. Chem. Phys. \textbf{148%
}, 204103 (2018).

\bibitem{28} S. Datta, \textit{Quantum Transport: Atom to Transistor},
Cambridge University Press, Cambridge (2005).

\bibitem{29} V. A. Sablikov and B. S. Shchamkhalova, Phys. Rev. B \textbf{58}%
, 13847 (1998).

\bibitem{30} V. A. Sablikov, S. V. Polyakov, M. B\"{u}ttiker, Phys. Rev. B
\textbf{61}, 13763 (2000).

\bibitem{31} J. P. Perdew, J. A. Chevary, S. H. Vosko, K. A. Jackson, M. R.
Pederson, D. J. Singh, and C. Fiolhais, Phys. Rev. B \textbf{46}, 6671
(1992).

\bibitem{32} C. S. Li, L. H. Wan, Y. D. Wei, and J. Wang, Nanotechnology
\textbf{19}, 155401 (2008).

\bibitem{33} L. Zhang, Y. X. Xing, and J. Wang, Phys. Rev. B \textbf{86},
155438 (2012).

\bibitem{34} D. R. Hamann, M. Schl\"{u}ter, and C. Chiang, Phys. Rev. Lett.
\textbf{43}, 1494 (1979).

\bibitem{35} M. C. Payne, M. P. Teter, D. C. Allan, T. A. Arias, and J. D.
Joannopoulos, Rev. Mod. Phys. \textbf{64}, 1045 (1992).

\bibitem{36} R. van Leeuwen, Phys. Rev. Lett. \textbf{82}, 3863 (1999).

\bibitem{37} M. Ruggenthaler and D. Bauer, Phys. Rev. A \textbf{80}, 052502
(2009).

\bibitem{45} Y. Kurzweil and R. Baer, J. Chem. Phys. \textbf{121}, 8731 (2004).

\bibitem{45-1} S. K. Ghosh and A. K. Dhara, Phys. Rev. A \textbf{38}, 1149 (1988).

\bibitem{45-2} I. V. Tokatly, Phys. Rev. B \textbf{71}, 165105 (2005).

\bibitem{45-3} G. Vignale and W. Kohn, Phys. Rev. Lett. \textbf{77}, 2037 (1996).

\bibitem{38} T. B. Boykin, Am. J. Phys. \textbf{68}, 665 (2000).

\bibitem{39} L. Kleinman and D. M. Bylander, Phys. Rev. Lett. \textbf{48},
1425 (1982).

\bibitem{40} S. W. Gao and Z. Yuan, Phys. Rev. B \textbf{72}, 121406(R)
(2005).

\bibitem{41} J. C. Phillips and L. Kleinman, Phys. Rev. \textbf{116}, 287
(1959).

\bibitem{42} Morrel H. Cohen and V. Heine, Phys. Rev. \textbf{122}, 1821
(1961).

\bibitem{43} Neil W. Ashcroft and N. David Mermin, \textit{Solid State
Physics}, Harcourt, Orlando (1976).
\end{thebibliography}
\end{document}